\documentclass[%
 reprint,
nofootinbib,
 amsmath,amssymb,
 aps,
 prx,
]{revtex4-2}

\usepackage{graphicx}
\usepackage{dcolumn}
\usepackage{bm}
\usepackage{xcolor}
\usepackage{makecell}
\usepackage{hyperref}
\usepackage{cellspace}
\begin{document}

\preprint{APS/123-QED}
\title{Confronting cosmic shear astrophysical uncertainties: DES Year 3 revisited} 

\author{Leah Bigwood,$^{1\ast}$ Jamie McCullough,$^{2}$ Jared Siegel$^{2}$, Alexandra Amon$^{2}$, George Efstathiou$^{1}$\\ David Sanchez-Cid$^{3}$, Elisa Legnani$^{4}$, Daniel Gruen$^{5,6}$, Jonathan Blazek$^{7}$, Cyrille Doux$^{8}$, Aurelio Carnero Rosell$^{9}$, Marco Gatti$^{10}$, Eric Huff$^{11}$, Niall MacCrann$^{12, 1}$, Anna Porredon$^{13}$, Judit Prat Marti$^{14}$, Marcelle Soares dos Santos$^{3}$, Justin Myles$^{2}$, Simon Samuroff$^{4}$, Masaya Yamamoto$^{2}$, Boyan Yin$^{15}$, Joe Zuntz$^{16}$ \& \textit{The Dark Energy Survey}\\
\vspace{2pt}
\small{\textit{(Affiliations can be found after the acknowledgements)}}\\
{\small{$^\ast$lmb224@cam.ac.uk}}
}




\begin{abstract}
Cosmology from weak gravitational lensing has been limited by astrophysical uncertainties in baryonic feedback and intrinsic alignments. By calibrating these effects using external data, we recover non-linear information, achieving a 2\% constraint on the clustering amplitude,  $S_8$, resulting in a factor of two improvement on the $\Lambda$CDM constraints relative to the fiducial Dark Energy Survey Year 3 model. The posterior, $S_8=0.832^{+0.013}_{-0.017}$, shifts by $1.5\sigma$ to higher values, in closer agreement with the cosmic microwave background result for the standard six-parameter $\Lambda$CDM cosmology. Our approach uses a star-forming `blue' galaxy sample with intrinsic alignment model parameters calibrated by direct spectroscopic measurements, together with a baryonic feedback model informed by observations of X-ray gas fractions and kinematic Sunyaev–Zel’dovich effect profiles that span a wide range in halo mass and redshift. Our results provide a blueprint for next-generation surveys: leveraging galaxy properties to control intrinsic alignments and external gas probes to calibrate feedback, unlocking a substantial improvement in the precision of  weak lensing surveys.
\end{abstract}

\maketitle


\section{\label{sec:level1}Introduction}

Weak gravitational lensing induces tiny distortions in the observed shapes of distant galaxies, due to the  gravitational influence of intervening cosmic structure. 
It offers an important test of gravity, dark matter and dark energy, probing the Universe across a wide range of redshifts and scales \cite{Weinberg_probes}. Weak lensing surveys have grown to measuring over 100 million galaxies, but the full cosmological power of these data remain untapped.  Cosmological inference 
is limited primarily by our ability to model two astrophysical effects \citep{amon_2022,KiDSDES}: (1) baryon feedback, the redistribution of gas within and beyond halos caused by active galactic nuclei and supernovae \citep[e.g.][]{vanDaalen2011}; and (2) intrinsic alignments (IA), the extent to which galaxy shapes align with their surrounding tidal gravitational field \citep[e.g.][]{TroxelIA}. In anticipation of unprecedented statistical power from, e.g., the Vera Rubin Observatory's Legacy Survey of Space and Time
\citep[LSST;][]{SRD-LSST}, ESA’s Euclid mission
\citep{euclid_overview}, and the Nancy Grace Roman Telescope
\citep{Roman2025}, there is pressing need for these two obstacles to be overcome. This paper presents a path forward.

Traditional modelling of baryonic feedback and intrinsic alignment introduces significant systematic uncertainties and potential biases to cosmic shear analyses. 
Baryon feedback suppresses the non-linear matter power spectrum relative to a dark matter only prediction \citep{Chisari2019}.
Without precise predictions for the impact of baryon feedback, weak lensing analyses typically adopt conservative scale cuts \citep[e.g.,][]{SeccoSamuroff_2022,amon_2022} - excluding the small-scale modes most sensitive to baryonic physics, but also missing valuable cosmological information.
Even when small scales are included, 
flexible feedback models \citep[e.g.,][]{Schneider:2015, Schneider2025, BACCO, Salcido2023,derose2025} fail to recover additional cosmological power without external calibration \citep[e.g.,][]{Bigwood2024, Decade2025, Arico2023,Schneider2022}.
This arises because there is a strong degeneracy between with the inferred strength of feedback and the lensing parameter, $S_8 = \sigma_8 \sqrt{\Omega_{\rm m}/0.3}$\footnote{Here $\sigma_8$ is the root-mean-square linear amplitude of the matter fluctuation spectrum in spheres of radius $8,h^{-1},\mathrm{Mpc}$, $h$ is the Hubble parameter in units of $100,\mathrm{km,s^{-1},Mpc^{-1}}$, and $\Omega_{\rm m}$ is the present-day matter density.} \citep{AAGPE2022, Preston2023}.

The inference of $S_8$ also depends on the assumed IA model \citep[e.g.][]{Lamman2024}. Two models have been widely used in the literature: the non-linear linear-alignment (NLA) and tidal alignment and tidal torquing (TATT) models. These models lead to results that typically differ by $>0.5\sigma$ \citep[e.g.,][]{KiDSDES, SeccoSamuroff_2022}.
Both models have been shown to be consistent with direct IA measurements, but only on larger scales ({$>$2-6 Mpc$/h$ and $k$$\sim$1-3$h/$Mpc). More complex models \citep[e.g.][]{Fortuna2021, Vlah2020,chen2024,Maion2024} have been developed, but these require additional model parameters that exacerbate the loss of cosmological precision.
In addition, in cosmic shear analyses, the entire galaxy sample is typically modelled by an IA model with a single set of parameters. In reality the intrinsic alignment of galaxies is known to depend on their properties \cite[e.g.,][]{Samuroff_2023,Johnston_2019,Siegel2025_IA, PAU_IA} such as type (e.g., star forming or quenched) and mass, which vary strongly with redshift within a typical weak lensing sample.

Cosmological hydrodynamical simulations provide predictions for the suppression of the matter power spectrum caused by baryon feedback \citep{flamingo, McCarthy2017, Dave2019, Hernandez-Aguayo2022, XFABLE}. However, these simulations depend on subgrid implementations of physics that cannot be modelled \textit{ab initio},
limiting their use to directly inform weak lensing analyses. 

Observational constraints on baryon feedback have improved rapidly in recent years. A number of new measurements indicate that the effects of feedback on the power spectrum are stronger than previously thought. These include
X-ray gas fraction observations from the eROSITA-DE survey \citep{Bulbul2024, Popesso2024, Kovac2025,Siegel:2025_kSZ}, 
kinetic Sunyaev–Zel’dovich (kSZ) effect measurements from the Atacama Cosmology Telescope (ACT) and the Sloan Digital Sky Survey (SDSS)/Dark Energy Spectroscopic Instrument (DESI) \citep{Schaan:kSZ, Bigwood2024,Hadzhiyska2024,McCarthy_kSZWL,ReidkSZ,Siegel:2025_kSZ,Bigwood_kSZsims}, diffuse X-ray emission \citep{Ferreira2024, LaPosta2024}, the thermal SZ effect \citep{TroestertSZ, LaPosta2024,Pandey25,Dalal25}, and fast radio bursts (FRBs) \citep{Reischke2025}.

Each observational probe is typically sensitive to only a narrow range of halo masses and redshifts, whereas weak lensing observations (without imposing scale cuts) are sensitive to a much broader range in both mass and redshift.  The complementarity of current X-ray and kSZ data can be used to overcome this limitation. For example, the measurements analyzed by \cite{Siegel:2025_kSZ} span a wide range of halo mass and redshift, $13<M_{500} [\mathrm{M_{\odot}}]<14.5$ and $0<z<1$ \cite{Siegel:2025_kSZ}, sufficient for the range that impacts full-scale weak lensing observations \cite[e.g.][their Fig.~9]{LucieSmith2025}.

Driven by massively multiplexed spectroscopic instruments, like DESI, and narrow-band imaging surveys, the growing landscape of IA direct measurements has revealed strong dependencies on galaxy type, luminosity, and colour \citep{Johnston_2019,Fortuna2021,Samuroff_2023, PAU_IA, Georgiou2025, Siegel2025_IA, Fortuna_2025}.
Red, quenched ellipticals show pronounced alignments, particularly in dense environments, whereas blue, star-forming galaxies exhibit no statistically significant alignment \citep{Samuroff_2023, Siegel2025_IA, Johnston_2019}.
The most precise measurements to date find no evidence of IA of blue galaxies between $0<z<1.5$ \citep{Siegel2025_IA}.
Recent analyses have begun to exploit this property in lensing studies, splitting samples by galaxy color or spectral type to isolate populations with minimal IA contamination \citep{des_y1_colorsplit, mccullough:2024_blue, wright2025, St_lzner_2025}.
These results suggest that informed sample selection, guided by spectroscopic measurements, offers a path to mitigate IA systematics in current and future weak lensing surveys.

This paper describes a cosmic shear analysis that combines constraints on baryonic feedback and direct measurements of IA to maximize the cosmological information. We adopt the baryon feedback model constraints of \cite{Siegel2025_BC}, which use galaxy-galaxy lensing characterized \citep{Siegel:2025_kSZ} X-ray gas fraction \citep{Bulbul2024} and kinematic Sunyaev–Zel’dovich (kSZ) \citep{Schaan:kSZ, ReidkSZ} measurements spanning a broad range of redshift and halo masses.
For intrinsic alignments, we build upon the color-based sample selection of \citep{mccullough:2024_blue} that restricts the lensing analysis to star-forming `blue' galaxies, roughly 70\% of the Dark Energy Survey Year 3 (DES Y3) weak lensing sample. We incorporate direct IA measurements for such a color selection of galaxies across redshift \citep{Siegel2025_IA, PAU_IA, Samuroff_2023} to inform a tailored-IA model (McCullough, Siegel et al. in prep.). We reanalyse the color-selected DES cosmic shear data using this combined strategy
and demonstrate the significant improvement in cosmological precision previously lost to astrophysical uncertainty. 

The analysis presented in this Letter provides a strategy for next-generation cosmic shear cosmology that incorporates data-driven models to exploit the full scale range of the measurement. 

\vspace{-0.5cm}
\section{\label{sec:level2} Cosmic shear data}
\vspace{-0.3cm}
The Dark Energy Survey (DES) observed in the \textit{grizy} bands on the Blanco 4-meter telescope at Cerro Tololo Inter-American Observatory. Between 2013-2019, DES collected imaging over 5000 deg$^2$ of the southern sky.  We revisit the Y3 cosmic shear analysis \cite{amon_2022,SeccoSamuroff_2022}, which used \textit{riz} \textsc{metacalibration} \cite{Sheldon_2017} shape and color-magnitude measurements \cite{Gatti:2021}.  We make use of a blue, star-forming source galaxy sample of 65,066,594 galaxies.  The selection, shear and redshift calibration was performed in \cite{mccullough:2024_blue}, using the methodology of \cite{Myles:2021,MacCrann:2022}. 

\vspace{-0.5cm}
\section{Astrophysical modelling}
\vspace{-0.2cm}

\begin{figure*}[ht!]
    \centering
        \includegraphics[width=0.8\textwidth]{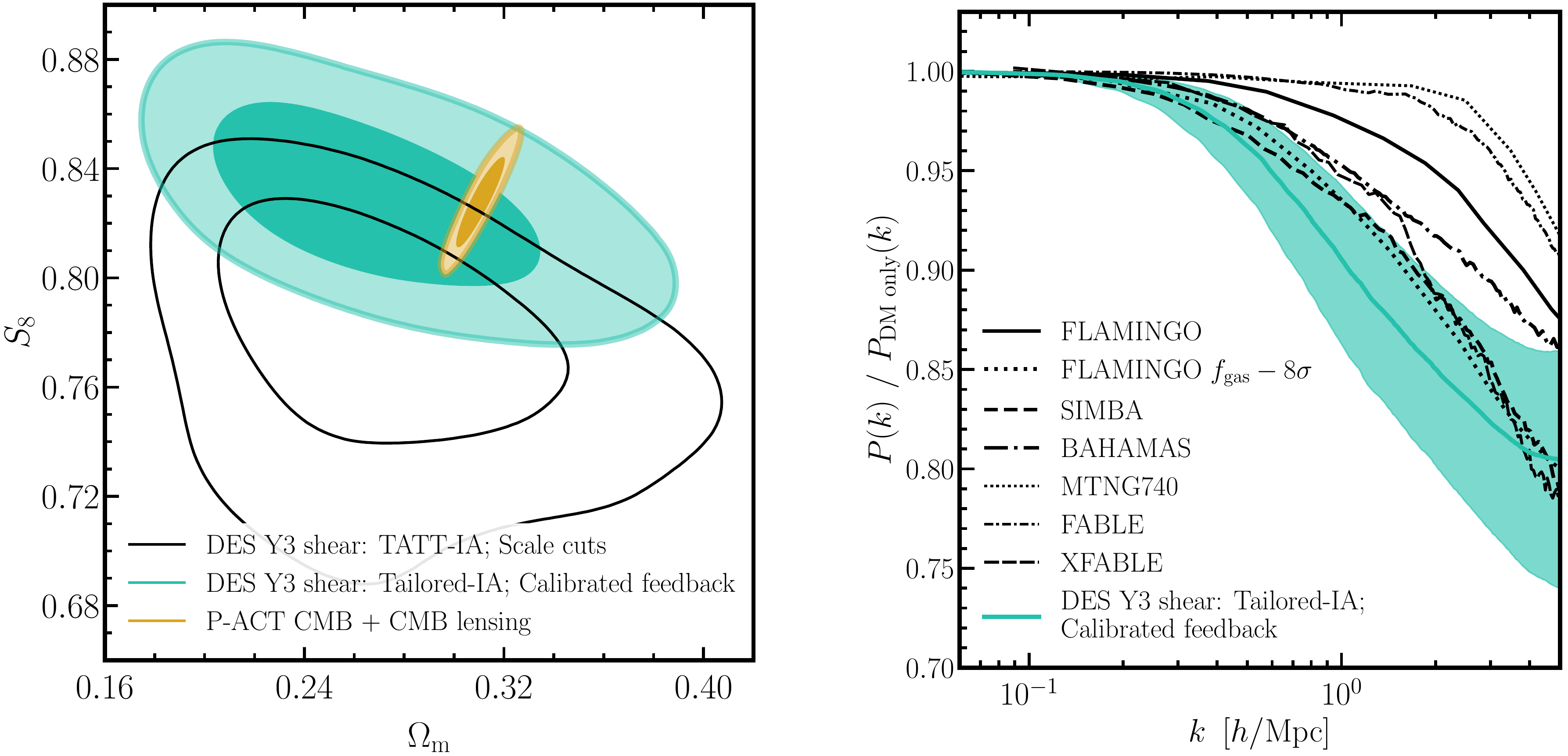}
        \vspace{-0.4cm}
    \caption{Marginalised $\Omega_{\rm m}-S_8$ posteriors (\textit{left}) and the suppression of the matter power spectrum due to baryonic feedback (\textit{right})
    constrained using DES Y3 reanalysed with calibrated feedback and tailored IA (teal).
    \textit{Left: } Our result shows a factor of $\times$2 improvement in $S_8$ uncertainty compared to the fiducial DES Y3 model (black; all galaxies, using scale cuts to mitigate baryonic feedback and the TATT IA model) and is in better agreement with the CMB $\Lambda$CDM constraint measured by ACT+Planck TTTEEE+lensing likelihood \citep{ACTDR6,ACT-ext}.  \textit{Right: }  By retaining the full scale extent of the cosmic shear measurements, we constrain the matter power spectrum suppression, 
    which, dominated by our informed prior, favors stronger feedback compared to  hydrodynamical simulations; FLAMINGO \& FLAMINGO $f_{\mathrm{gas}}-8\sigma$\citep{Schaye2023}, SIMBA \citep{Dave2019}, BAHAMAS \citep{McCarthy2017}, MTNG740 \citep{Pakmor2023} and (X)FABLE \citep{Henden2018, XFABLE}. }
    \label{fig:results}
\end{figure*}

\vspace{-0.1cm}
\subsection{Calibrated baryonic feedback model}\label{sec:BFmodel}
\vspace{-0.3cm}
In order to robustly use astrophysical data to inform a weak lensing baryon feedback model, we require: 
\vspace{-0.2cm}
\begin{enumerate}
\itemsep=0em
\item The model passes injection recovery tests to demonstrate that it can map between astrophysical observables and the suppression of the matter power spectrum to reproduce the results from a range of hydrodynamical simulations \citep{Schneider2019, Giri2021, Siegel2025_BC}. 
\item The astrophysical data are directly studied in hydrodynamical simulations allowing cross-checks against results from parametric models of feedback processes \citep{McCarthy_kSZWL, Siegel:2025_kSZ, Bigwood_kSZsims}. 
\item  The sensitivity of the astrophysical data as a function of halo mass is well characterized, e.g. via calibration with galaxy-galaxy lensing, \citep{Siegel:2025_kSZ} in order to infer the strength of feedback.

\item The astrophysical data \citep{Schaan:kSZ, ReidkSZ, Bulbul2024} span the range of halo mass and redshift that the cosmic shear measurements are sensitive to \citep{LucieSmith2025,Siegel:2025_kSZ}. 
\end{enumerate}
We model the impact of baryonic feedback on the non-linear matter power spectrum using the \textit{baryonification} \texttt{BCEmu} model \citep{Schneider2019, Giri2021}, which perturbatively shifts particles in N-body simulations to mimic the effect of baryonic feedback on the total matter distribution.  Dark matter-only profiles are transformed spherically symmetrically to a total matter profile that comprises collisionless matter, gas, and central galaxy components.  We adopt this model for its demonstrated success in mapping between gas observables and the suppression of the matter power spectrum 
to reproduce a range of simulations \citep{Schneider2019, Giri2021, Siegel2025_BC}.  

We adopt the calibrated baryonification model from \cite{Siegel2025_BC}, which use 
eROSITA X-ray gas fractions \cite{Bulbul2024} and ACT+SDSS+DESI kSZ profiles \cite{Schaan:kSZ, ReidkSZ}, with mass calibration for both datasets using galaxy-galaxy lensing \cite{Siegel:2025_kSZ}. Together, these data span a wide range of halo masses and redshifts, $13<M_{500} [\mathrm{M_{\odot}}]<14.5$ and $0<z<1$ \cite{Siegel:2025_kSZ}, overlapping the sensitivity of weak lensing observations \cite{LucieSmith2025}. In \cite{Siegel2025_BC}, these data are modelled to constrain the suppression of the matter power-spectrum, providing correlated
priors on the feedback model parameters (App.~\ref{app:priors}). Their analysis incorporates several conservative choices for uncertainties in the astrophysical measurements: \\
(1) The statistical uncertainty on the galaxy-galaxy lensing estimate for the halo mass; 
\\
(2) A 10\% systematic uncertainty on the kSZ velocity reconstruction, doubled from the $<5$\% reported by \cite{Hadzhiyska2024velocities}; \\
(3) A $10\%$ systematic uncertainty on the mean gas fraction of eROSITA clusters to reflect the $15\%$ flux offset 
with \textit{Chandra} on common sources \citep[][]{Bulbul2024}, reported as $<1\%$.
In App.~\ref{app:xraykszchoice}, we consider analyses with a calibrated model informed by only the kSZ measurements, only the eROSITA gas fractions and by HSC-XXL gas fractions.

\vspace{-0.5cm}
\subsection{Tailored intrinsic alignment model}\label{sec:tailorIA}
\vspace{-0.4cm}
Modelling IA introduces a source of uncertainty and potential bias in cosmological constraints inferred from cosmic shear. 
However, not all galaxy samples are expected to show IA.
Direct measurements consistently find that blue, star forming galaxies are insignificantly, hence only weakly, aligned with their local environments \citep{GAMA_IA,Siegel2025_IA,Samuroff_2023, KiDSBrightIA, PAU_IA}. 
Recent measurements with DESI~DR1 spectroscopy show no evidence of IA of blue galaxies out to at least $z=1.5$ \citep{Siegel2025_IA}.
Cosmological constraining power can therefore be improved by restricting cosmic shear analyses to galaxy samples that are free of alignment \citep{mccullough:2024_blue}.
In this work, we consider the bespoke DES~Y3 \textit{blue} shear catalogue  \citep{mccullough:2024_blue}, which consists of a pure selection of star forming galaxies.
For each tomographic bin, we impose a Gaussian prior on the amplitude of IA that is centred at zero with a width based on the precision of the latest direct measurements \cite{Samuroff_2023,Siegel2025_IA,PAU_IA}. 
This approach is described in more detail in App.~\ref{app:tailoredia} and is explored further in McCullough, Siegel et al. (in prep.).

\begin{table*}
\begin{tabular}{ccccccccccc}
\hline
\hline
Model & $S_{8, \mathrm{mean}}\ (\sigma_{S_8})$ & $\Omega_{{\rm m}, \mathrm{mean}}$& $S_{8}^{\mathrm{MAP}}$ &  $\Omega_{\rm m}^{\mathrm{MAP}}$ &$\chi_{\rm min}^2$ & $N_{\rm d}$ & $N_{\rm p}$ &$\chi_{\rm red}^2$ & $\Delta S_8$ & $\times\sigma_{S8}$ \\
\hline
Tailored-IA \& Calibrated Feedback & $ 0.832^{+0.013}_{-0.017}$ ($\pm 0.016$) & $0.252^{+0.017}_{-0.027}$ &0.849 & 0.236 & 379.8&400& 4.3& 1.0 & 0.1$\sigma$ & 2.0\\ 

Tailored-IA \& Un-calibrated Feedback & $ 0.820^{+0.016}_{-0.025} $ $(\pm0.022)$ & $ 0.264^{+0.023}_{-0.033} $  &0.804&0.271 &379.6&400& 5.0 &1.0&0.4$\sigma$ & 1.5\\ 
Tailored-IA \& Scale cuts & $ 0.803^{+0.016}_{-0.022} $ $(\pm 0.019)$ & $0.314^{+0.027}_{-0.045} $& 0.819 &0.302 & 258.2 & 273 & 3.9 & 1.0&1.1$\sigma$& 1.7 \\ 
TATT-IA \& Scale cuts (DES fid. choices) & $ 0.777^{+0.035}_{-0.024} $ ($\pm 0.032$)& $0.273^{+0.035}_{-0.052}$&  0.776 &0.308&293.1&273& 6.3 &1.1&1.5$\sigma$ & - \\ 

\hline
\end{tabular}
\vspace{-0.2cm}
\caption{Cosmological results (mean and 68\% confidence levels, standard deviation ($\sigma_{S_8}$) and \textit{maximum a posteriori} (MAP) values) with the goodness of fit, $\chi^2_{\rm red}=\chi^2_{\rm min}/(N_{\rm d}-N_{\rm p})$, defined in terms of the $\chi^2$ minimum value, $\chi^2_{\rm min}$, the number of data points analyzed, $N_{\rm d}$ and the effective number of parameters constrained, $N_{\rm p}$\footnote{These are computed similarly to \cite{SeccoSamuroff_2022,raverihu} with \textsc{cosmosis} \cite{Zuntz2015}, where $N_{\rm{p}} = N_{\rm{param}} - \mathrm{Tr}(C_{\rm{post}}C_{\rm{prior}}^{-1})$.}. We compare the $S_8$ value to the CMB result \citep{ACTDR6}, $\Delta S_8=(S_8^{\rm CMB}-S_8)/({\sigma_{S_8, {\rm CMB}}^2 + \sigma_{S_8}^2})^{0.5}$, 
and the improvement in the $S_8$ uncertainty, $\times\sigma_{S8}$, compared to the result using the fiducial DES Y3 choices (bottom row).
}
\label{tab:results}
\vspace{-0.2cm}
\end{table*}

\vspace{-0.2cm}
\section{Joint constraints on Cosmology \& astrophysics}
\vspace{-0.2cm}
This section presents the constraints on cosmological and astrophysical parameters and the suppression of the matter power spectrum. To analyze the cosmic shear data, we generally follow the public DES Y3
cosmological inference pipeline \citep{Amon:2021, SeccoSamuroff_2022} within the \textsc{Cosmosis} framework \citep{Zuntz2015}, using the \textsc{polychord} sampler \citep{Handley2015a,handley2015b} with the optimal sampler settings \citep{Lemos2023,KiDSDES} (see App.~\ref{app:priors}).  We make the following changes to the analysis choices. We adopt uninformative priors on cosmological parameters, but assume a fixed neutrino mass, $m_\nu=0.06$eV for the heaviest neutrino mass eigenstate. 
For each redshift bin, we model the uncertainty in the mean redshift and the shear calibration for the blue sample following the prior choices of \citep{mccullough:2024_blue}.  
We compute the linear matter spectrum using \texttt{CAMB} \citep{Lewis2000} and follow \citep{KiDSDES} in modelling the non-linear correction with \texttt{HMcode2020} \citep{Mead2021}.  All prior choices are reported in Table~\ref{tab:priors}.

\textit{Cosmology:} The 2D marginalized constraints on the lensing amplitude $S_8$ and matter density $\Omega_{\rm m}$ are shown in teal in the left panel of Fig.~\ref{fig:results} and reported in Table \ref{tab:results}. The mean and 68\% credible intervals (\textit{maximum a posteriori}, MAP and standard deviation) values of $S_8$ are
\label{eqn:S8results}
\begin{equation}
  S_8 =\,\,\,  0.832^{+0.013}_{-0.017} \,\, (0.849 \pm 0.016)
\end{equation}
\noindent 
which has a 2\% uncertainty, a $\times$2 improvement compared to the result with the DES Y3 model of scale cuts and the TATT-IA model. We note that the posterior shifts by 1.5$\sigma$ to higher values of $S_8$ compared to DES Y3 fiducial choices.  We compare to the CMB constraint from ACT+\textit{Planck} \citep{ACTDR6} and find consistency within 0.1$\sigma$.  In App.~\ref{app:comparison} we compare our constraints to other cosmological analyses in the literature. 

The changes driving the improvement in precision relative to the DES Y3 analysis choices are shown in Fig.~\ref{fig:analysischoices}.  Retaining the fiducial DES Y3 baryonic mitigation of scale cuts, but replacing the TATT model with the tailored IA treatment (`Tailored-IA; Scale cuts'), yields a factor of 1.7 improvement in $S_8$ uncertainty.  Replacing the scale cuts with our X-ray- and kSZ-calibrated \texttt{BCEmu} priors improves the constraint by a further factor of 1.2. 

\begin{figure}
    \centering
    \includegraphics[width=1\linewidth]{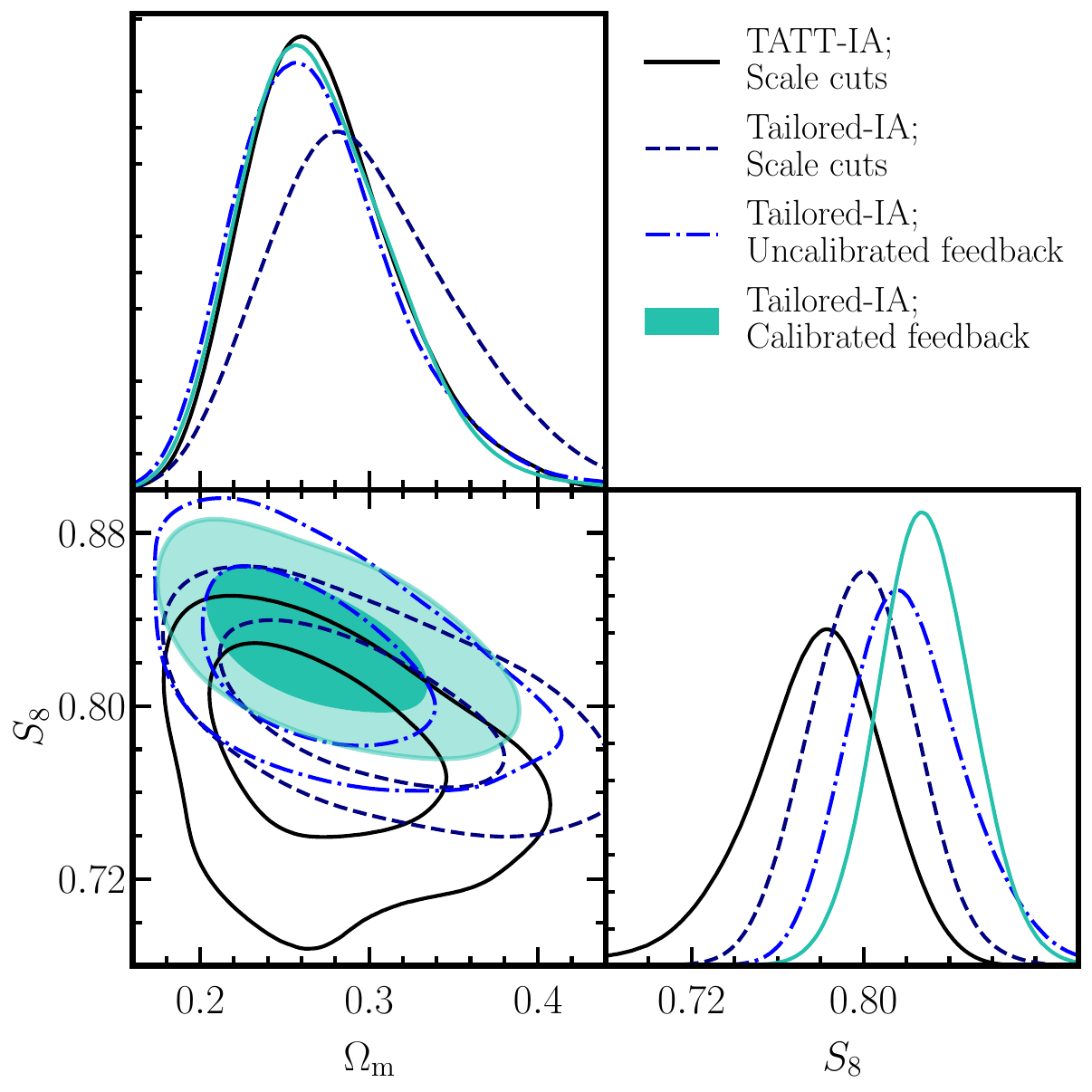}
    \vspace{-0.7cm}
    \caption{Marginalized $\Omega_{\rm m}-S_8$ posteriors constrained using DES Y3, showing the impact of each model improvement compared to the fiducial DES choices (TATT IA model with scale cuts, black): Tailored-IA with scale cuts (navy dashed); with an uninformed feedback model without scale cuts (blue, dot-dashed); and with calibrated feedback (teal). }
    \vspace{-0.4cm}
    \label{fig:analysischoices}
\end{figure}



\textit{Astrophysics:}
We constrain the suppression of the matter power spectrum to 4.3\% precision at $k=1 \ h/\mathrm{Mpc}$. At $k = 2$ (and $k = 0.5$) $h/\mathrm{Mpc}$, 
we find a stronger suppression of the matter power spectrum than predicted by FLAMINGO by $2.2\sigma$ ($1.6\sigma$), SIMBA by $0.8\sigma$ ($0.3\sigma$), BAHAMAS by $1.5\sigma$ ($1.0\sigma$), MTNG740 by $3.1\sigma$ ($1.8\sigma$) and FABLE by $2.9\sigma$ ($1.8\sigma$).  These findings are broadly in agreement with recent studies informed by observational gas probes \citep{Schneider2022,Bigwood2024,Dalal2025, Pandey25,LaPosta2024, Kovac2025, Grandis:2024,fabbian2025}. 


\vspace{-0.3cm}
\section{Outlook}
\vspace{-0.2cm}
Cosmological inference for weak lensing studies is limited by systematics in the modelling of baryon feedback and intrinsic alignments. Underestimating feedback \citep{Amon2022} or inaccurately modelling IA \citep{Krause} biases cosmological parameters, while flexible but uncalibrated models sacrifice constraining power \citep{Amon:2021, SeccoSamuroff_2022,KiDSDES}. Our data-driven approach -- combining constraints on the matter power spectrum with direct measurements of IA -- demonstrates that both obstacles can be overcome with current data.

This analysis represents a shift from mitigating astrophysical effects by excluding data on small scales to measuring and calibrating them. 
By linking the gas physics traced by SZ and X-ray observations with galaxy-shape alignments, we recover non-linear information and achieve a factor-of-two improvement in $S_8$ precision, bringing weak lensing results into closer agreement with CMB constraints in $\Lambda$CDM. 
By using all scales of the cosmic shear data, we also constrain the suppression of the matter power spectrum and the gas fraction - halo mass relation, finding that the data prefers a feedback scenario that is stronger than predicted by simulations.

The strategy presented here provides a blueprint for next-generation weak lensing surveys \citep{LSST, Euclid, Roman} and an opportunity to unlock their full statistical power. X-ray and kSZ measurements of baryon feedback will improve in precision via observations from eROSITA, DESI and Simons Observatory and will span a wider range in halo mass and redshift. Direct measurements of intrinsic alignments will allow for an optimised selection of the weak lensing sample, potentially including red galaxies with well-measured IA. Our approach to calibrating the astrophysical model has natural extensions with measurements of thermal SZ and cluster relations \citep{TroestertSZ,LaPosta2024,Dalal25,Pandey25}, diffuse X-ray \citep{Ferreira2024} and FRB dispersion measures \citep{KrittiFRB,Reischke2025}, provided that these data are well characterized (see Section~\ref{sec:BFmodel}).
In the future, we anticipate using this strategy with improved models \citep[e.g.][]{Schneider2025,Pandey2025, Shavelle2025,lague}.

This approach can transform weak lensing from a systematics-limited probe to unlocking its full potential to test the standard model and the galaxy formation.

%

\section*{Acknowledgments}
LB acknowledges support from the Science and Technology Facilities Council (STFC).  JS acknowledges support by the National Science Foundation Graduate Research Fellowship Program under Grant DGE-2039656. 
Any opinions, findings, and conclusions or recommendations expressed in this material are those of the author(s) and do not necessarily reflect the views of the National Science Foundation. AA is grateful for support from a Sloan Fellowship. DG was supported by the Deutsche Forschungsgemeinschaft (DFG, German Research Foundation) under Germany's Excellence Strategy – EXC-2094 – 390783311.

This project used public archival data from the Dark Energy Survey (DES).

\section*{Affiliations}
\footnotesize{
{$^{1}$Institute of Astronomy, University of Cambridge, Madingley Road, Cambridge CB3 0HA, UK}\\
{$^{2}$Department of Astrophysical Sciences, Princeton University, 4 Ivy Lane, Princeton, NJ 08544, USA}\\
{$^{3}$Physik-Institut, University of Zürich, Winterthurerstrasse 190,
 CH-8057 Zürich, Switzerland}\\
 {$^4$}Institut de Física d’Altes Energies (IFAE), The Barcelona Institute of Science \& Technology, Campus UAB, 08193 Bellaterra (Barcelona) Spain\\
{$^{5}$University Observatory, Faculty of Physics, Ludwig-Maximilians-Universität, Scheinerstr. 1, 81679 Munich, Germany}\\
{$^{6}$Excellence Cluster ORIGINS, Boltzmannstr. 2, 85748 Garching, Germany}\\
{$^{7}$Department of Physics, Northeastern University, Boston, MA 02115, USA}\\
{$^{8}$Université Grenoble Alpes, CNRS, LPSC-IN2P3, 38000 Grenoble, France}\\
{$^{9}$Instituto de Astrofisica de Canarias, E-38205 La Laguna, Tenerife, Spain}\\
{$^{10}$Kavli Institute for Cosmological Physics, University of Chicago, Chicago, IL 60637, USA}\\
{$^{11}$Jet Propulsion Laboratory, California Institute of Technology}\\
{$^{12}$DAMTP, Centre for Mathematical Sciences, University of Cambridge, Wilberforce Road, Cambridge CB3 OWA, UK }\\
{$^{13}$Centro de Investigaciones Energ\'eticas, Medioambientales y Tecnol\'ogicas (CIEMAT), Madrid, Spain}\\
{$^{14}$Nordita, KTH Royal Institute of Technology and Stockholm University, Hannes Alfv\'ens v\"ag 12, SE-10691 Stockholm, Sweden}\\
{$^{15}$Department of Physics, Duke University Durham, NC 27708, USA}\\
{$^{16}$Institute for Astronomy, University of Edinburgh, Royal Observatory, Blackford Hill, Edinburgh, EH9 3HJ, U.K}\\
}
\normalsize
\appendix

\section{Priors, Samplers \& Posteriors}\label{app:priors}

\begin{table}
    \caption{The cosmological, observational and astrophysical parameters and the priors that we adopt. The `uninformed' priors for baryonic feedback parameters correspond to the fiducial \texttt{BCEmu} priors constructed to span a conservative range hydrodynamical simulations \citep{Schneider2019}; the `calibrated' are informed by X-ray and kSZ measurements \citep{Siegel2025_BC}.  In practice, we exploit the covariance between BCEmu parameters constrained by data in \citep{Siegel2025_BC}, and use correlated priors bounded by the uninformed BCEmu limits.  Here, for simplicity, we show the corresponding 1D Gaussian marginal priors.  `Calibrated' IA priors are derived per redshift bin from direct measurements of star-forming galaxies. For flat priors, the range is denoted by square brackets, while Gaussian priors are denoted as (mean, 1$\sigma$ width).}
    \label{tab:priors}
\begin{center}
\begin{tabular}{ccc}
\hline
\hline
Parameter & Uninformed  & Calibrated \tabularnewline
\hline 
\bf{Cosmological} \tabularnewline
$\Omega_{\rm m}$, Matter density &  [0.1, 0.9] & See Fig.~\ref{fig:fbprior} \tabularnewline
 $\Omega_{\rm b}$, Baryon density &  [0.03, 0.07] & - \tabularnewline
$10^{-9}A_{\rm s}$, Sc. spec. amp.  & [0.5, 5.0] & -\tabularnewline
$h$, Hubble parameter  &  [0.55, 0.91] & - \tabularnewline
$n_{\rm s}$, Spectral index  &  [0.87,1.07] & - \tabularnewline
\hline 
\bf{Observational} \tabularnewline
$\Delta z^1$, Redshift 1 & - & ( 0.0, 0.018 ) \tabularnewline
$\Delta z^2$, Redshift 2 & - & ( 0.0, 0.015 ) \tabularnewline
$\Delta z^3$, Redshift 3 & - & ( 0.0, 0.011 ) \tabularnewline
$\Delta z^4$, Redshift 4  & - & ( 0.0, 0.017 ) \tabularnewline
$m^1$, Shear calibration 1 & - & ( -0.006, 0.009 )\tabularnewline
$m^2$, Shear calibration 2 & - & ( -0.020, 0.008 )\tabularnewline
 $m^3$, Shear calibration 3 & - & ( -0.024, 0.008 )\tabularnewline
$m^4$, Shear calibration 4 & - & ( -0.037, 0.008 )\tabularnewline
\hline 
\bf{Astrophysical} \tabularnewline
$a_1$, IA amplitude 1    &  [-3,3]  &  ( 0, 0.66 ) \tabularnewline
$a_2$, IA amplitude 2    &  [-3,3]  &  ( 0, 0.77 ) \tabularnewline
$a_3$, IA amplitude 3    &  [-3,3]  &  ( 0, 0.68 ) \tabularnewline
$a_4$, IA amplitude 4    &  [-3,3]  &  ( 0, 0.86 ) \tabularnewline

\hline 

$\log_{10}M_{\rm c}$ & [11, 15] & (13.60, 0.17)  \tabularnewline
$\theta_{\rm ej}$ & [2, 8] & ( 5.85, 1.20)  \tabularnewline
$\mu$  & [0, 2]  & (0.92,  0.14) \tabularnewline
$\gamma$   &  [1,4] & (2.27, 0.77) \tabularnewline
$\delta$ & [3,11] & (7.71, 1.70) \tabularnewline
$\eta_{\delta}$  & [0.05, 0.04] & (0.20, 0.05) \tabularnewline
$\eta$ &  [0.05.4] & (0.11, 0.04)  \tabularnewline
\hline 

\end{tabular}
\end{center}
\end{table}

Table~\ref{tab:priors} summarises the `uninformed' and `calibrated' priors adopted in our analysis for cosmological, observational, and astrophysical parameters.
We adopt wide cosmology priors for both the `uninformed' and `calibrated' priors. Figures~\ref{fig:IAparams} and \ref{fig:BCparams} show the priors and the posteriors on the IA and baryonic feedback model parameters constrained using DES data, analysed with and without the calibrated priors. Marginalizing the calibrated feedback prior with the uniformed cosmology prior significantly widens the prior on matter power suppression (Figure \ref{fig:BCparams}). These priors enter our cosmological analysis for which we use the \textsc{polychord} sampler settings in Table~\ref{tab:polychord} to compute the summary statistics reported.

\begin{table} 
    \caption{\textsc{polychord} sampler settings, slightly modified from the `high-resolution' recommended settings in \cite{Lemos2023} to allow for MAP estimation.}
     \label{tab:polychord}
    \begin{center}
    
    \begin{tabular}{c|c}
        \hline
        \hline
        Sampler Setting & Value \tabularnewline
        \hline
         \texttt{polychord.live\_points} & 500 \tabularnewline
         \texttt{polychord.tolerance} & 0.01 \tabularnewline
         \texttt{polychord.fast\_fraction} & 0.01 \tabularnewline
         \texttt{polychord.num\_repeats} & 60 \tabularnewline
         \texttt{polychord.boost\_posteriors} & 10.0 \tabularnewline
         \hline
    \end{tabular}
    
    \end{center}
    
\end{table}

\begin{figure}
    \centering
    \includegraphics[width=1\linewidth]{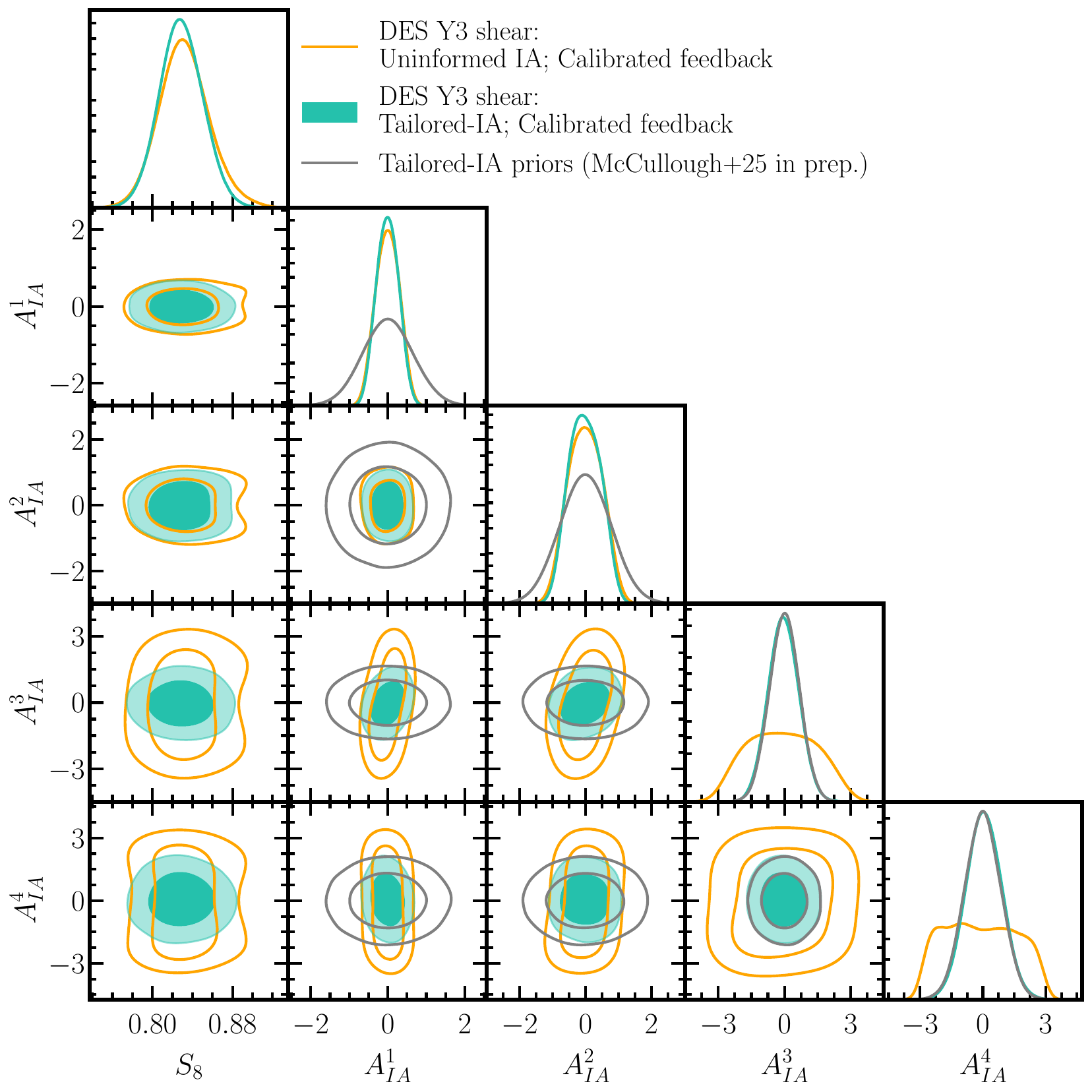}
    \caption{Marginalised  posteriors on $S_8$ and the IA amplitude in four tomographic bin, modelled using NLA.  We show constraints from the DES Y3 blue galaxy sample using tailored-IA and calibrated feedback, with the tailored-IA priors shown in grey (Section~\ref{sec:tailorIA}, McCullough, Siegel et al., in prep.). }
    \label{fig:IAparams}
\end{figure}

\begin{figure*}
    \centering
    \includegraphics[width=0.9\linewidth]{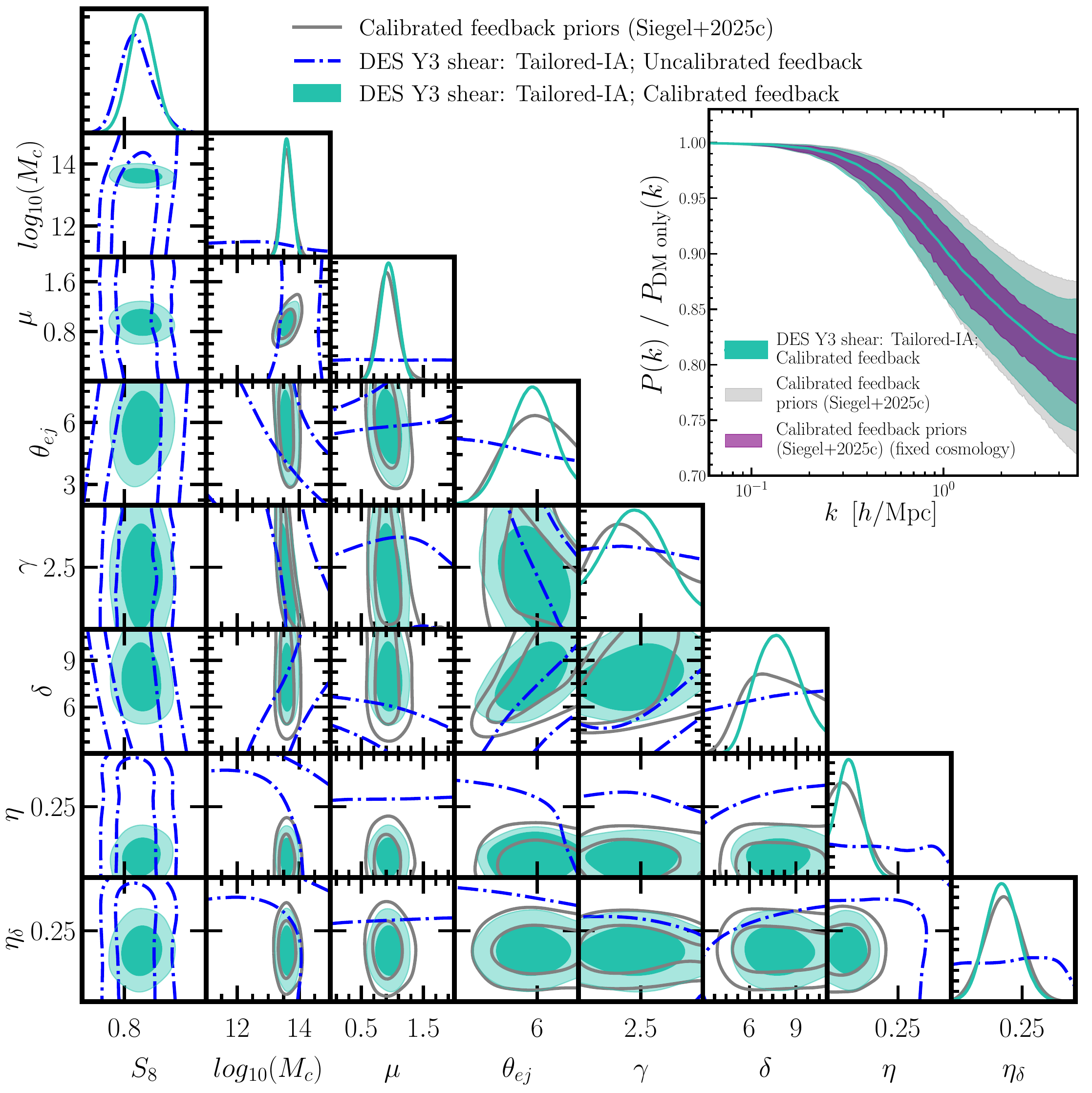}
    \caption{\textit{Left:} Marginalised posteriors on $S_8$ and BCEmu parameters constrained using the DES Y3 blue galaxy sample with uncalibrated (blue dash-dot) and calibrated (teal) baryonic feedback, with the calibrated feedback prior of \citep{Siegel2025_BC} shown in grey (Section~\ref{sec:BFmodel}, grey).  Both variants are analyzed with tailored-IA. \textit{Right:} Marginalised posterior on the suppression of the matter power spectrum due to baryon feedback with the calibrated prior (teal). The baryon feedback prior of \citep{Siegel2025_BC} (fixed cosmology) is shown in purple and the combination of the feedback prior and wide cosmology prior is shown in grey.}
    \label{fig:BCparams}
\end{figure*}

The \texttt{BCEmu} emulator \citep{Giri2021} depends on cosmology only through the baryon fraction,
$f_{\rm b}=\Omega_{\rm b}/\Omega_{\rm m}$, and was trained only over the range $0.1 < f_{\rm b} < 0.25$.
Because the emulator fails outside this interval, our analysis variants using \texttt{BCEmu} cannot
effectively explore the full cosmological prior $0.1 < \Omega_{\rm m} < 0.9$.

In Figure~\ref{fig:fbprior}, we show for our headline `Tailored-IA; Calibrated feedback' analysis the impact of the $f_b$ limits when drawing samples from the $\Omega_{\rm m}$ prior and
evaluating the likelihood, illustrating the regions where the likelihood evaluation fails (grey).  The resulting effective prior peaks around $\Omega_{\rm m} \sim 0.3$ and decreases such that
values $\Omega_{\rm m} \gtrsim 0.7$ are never sampled.

We also show the inferred constraints on $\Omega_{\rm m}$ from the same analysis evaluated on the data (teal). We find that the posterior lies well within the effective $\Omega_{\rm m}$ prior range, indicating that this limitation does not restrict our inference of $\Omega_{\rm m}$.

\begin{figure}
    \centering
    \includegraphics[width=0.7\linewidth]{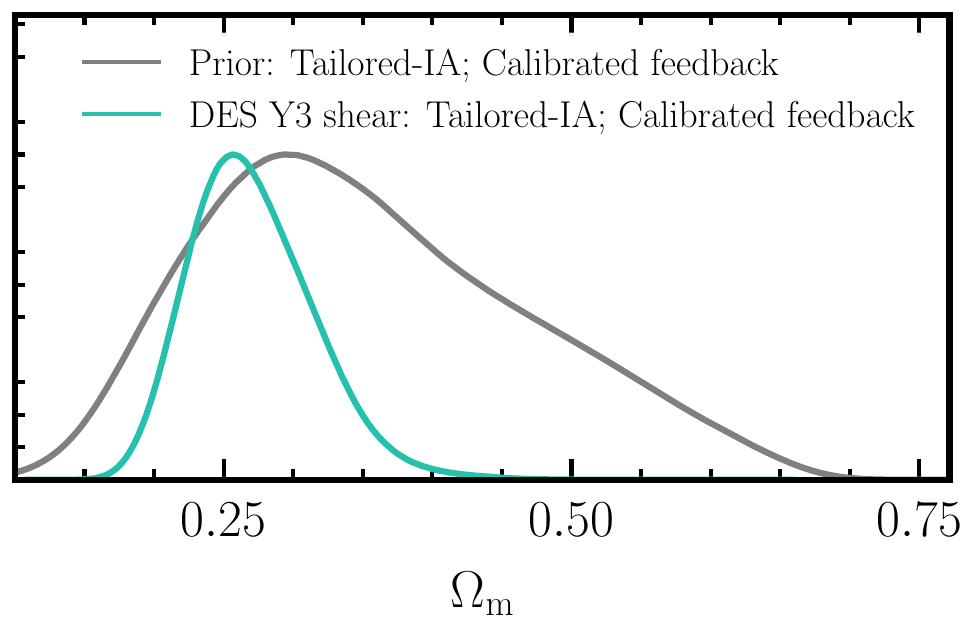}
    \caption{The effective prior on $\Omega_{\rm m}$ in the `Tailored-IA; Calibrated feedback' analysis, due to the \texttt{bcemu} emulator failing outside $0.1 < f_b < 0.25$ (grey).  The inferred $\Omega_{\rm m}$ posterior attained when evaluating on the data lies within the effective $\Omega_{\rm m}$ prior (teal).}
    \label{fig:fbprior}
\end{figure}

\section{Tailored-IA}\label{app:tailoredia}

The priors on galaxy-galaxy intrinsic alignment in this work are informed by direct measurements of blue, star-forming galaxy populations. With no significant alignment found in shape-position measurements for blue populations \cite[e.g.,][]{Siegel2025_IA, PAU_IA, Samuroff_2023, unions, Johnston_2019, Georgiou2025}, the most significant uncertainty to consider for this class of galaxy relates to the direct measurements themselves. These uncertainties are driven statistically by the number of available pairs of blue spectroscopic galaxies as a function of spatial separation -- which vastly decreases for high-redshift tracers like Emission Line Galaxies ($z > 1$). For this reason, we supply a prior on the amplitude of the IA, ($A_{\rm IA}$, as per the NLA model \cite{Bridle2007}) for each redshift bin of the blue shear galaxies that depends \textit{only} on the redshift distribution of that bin and the uncertainties on the null detections of the IA measurements.

\begin{figure}
    \centering
    \includegraphics[width=\columnwidth]{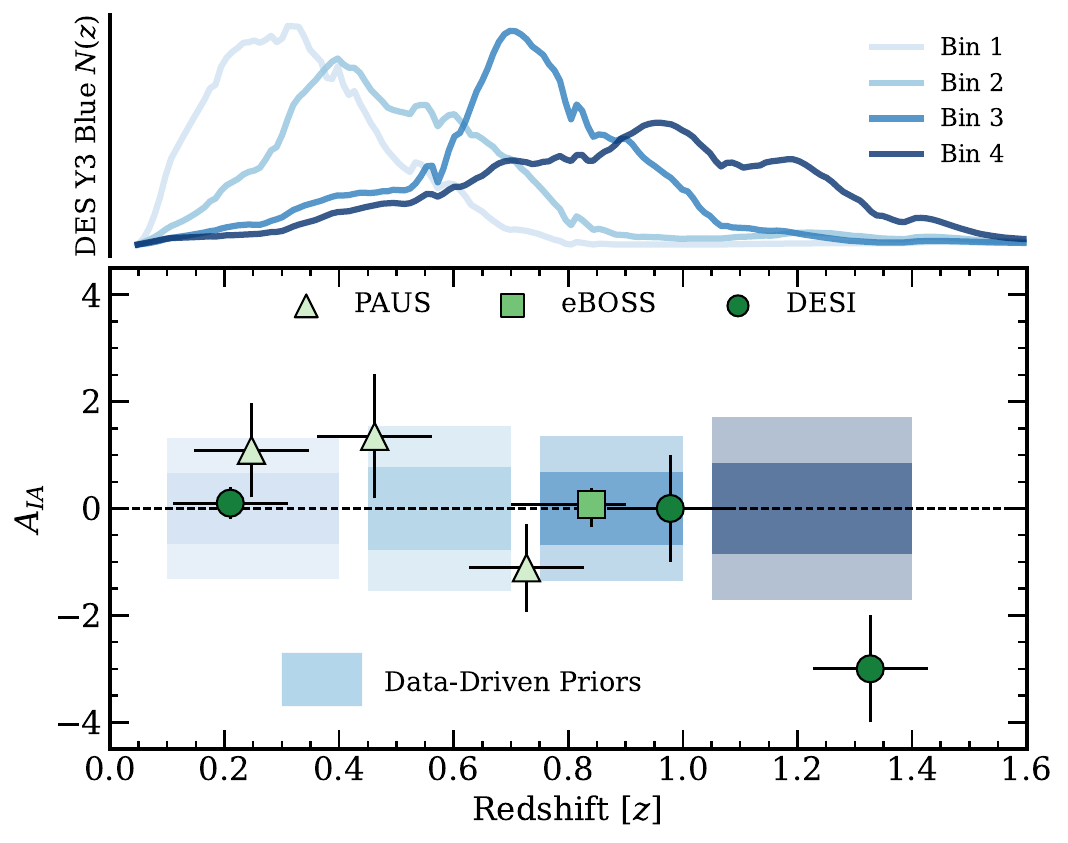}
    \caption{The construction of Tailored-IA priors on an IA amplitude as a function of redshift. \textit{Upper:} Redshift probability distributions for the blue galaxy selection from the DES Y3 weak lensing sample \citep{mccullough:2024_blue}. \textit{Lower:} Constraints on the IA amplitude, $A_{\rm IA}$, from direct IA measurements using blue star-forming galaxies from DESI \citep{Siegel2025_IA}, eBOSS \citep{Samuroff_2023} PAUS \citep{PAU_IA} as a function of their redshift. The priors are centred on zero, and their width reflects the uncertainty in the measurments.}
    \label{fig:tailoredia}
\end{figure}

As no measure of alignment in these galaxies exists with high-confidence, we center our per-bin priors at zero. We make use of the DESI Y1 direct measurements of blue-selected Bright Galaxy Survey and Emission Line Galaxies (split into low-, $z<1.0$, and high-redshift, $z>1.0$, samples respectively) \cite{Siegel2025_IA}, alongside a complementary (in redshift) Emission Line Galaxy sample from eBOSS \cite{Samuroff_2023}. Mid-to-low redshifts are well constrained by the narrowband photo-$z$ measurements from the PAUS survey \citep{PAU_IA}. We produce for each spectroscopic sample, $s$, in some calibration set of samples, $C$, a measure of the similarity in redshift to a given tomographic bin, $b$. This is the overlap coefficient, $F_{\rm s,b}$,
\begin{equation}\label{eqn:overlap}
    F_{\rm s,b} = \int \mathrm{min}[n_{\rm b}(z),n_{\rm s}(z)] dz,
\end{equation}
for normalized spectroscopic and tomographic redshift distributions $n_{\rm s}(z)$ and $n_{\rm b}(z)$, respectively. We primarily use this measure to obtain the uncalibrated fraction of photometric galaxies, where we impose a conservative prior for the portion of the tomographic bin with no direct measurements, so called uncalibrated, $u$. This uncalibrated fraction $F_{\rm u,b}$, can be derived similarly to Eqn.~\ref{eqn:overlap},
\begin{equation}
    F_{\rm u,b} = 1 - \int \mathrm{min}[n_{\rm b}(z), \mathrm{max}_{s\in C}\ n_{\rm s}(z)] dz,
\end{equation}
with the redshift-dependent uncalibrated component of the distribution written as
\begin{equation}\label{eqn:uncal}
    f_{\rm u,b}(z) = 1 - \frac{\mathrm{min}[n_{\rm b}(z), \mathrm{max}_{\rm s\in C}\ n_{\rm s}(z)]}{n_{\rm b}(z)} \; .
\end{equation}
For direct measurements that have overlapping redshift distributions, we can combine the measurements in that redshift slice with inverse-variance weights, and weighing their relative contributions to a given redshift. This gives us an expression for the combined measurement uncertainty in non-alignment as a function of redshift, $\Delta A_{\mathrm{m}}(z)$, as
\begin{equation}
    \Delta A_{\mathrm{m}}(z) = \sqrt\frac{\sum_s n_{\rm s}(z)}{\sum_s n_{\rm s}(z)/\Delta A_{\rm s}^2} \; .
\end{equation}
We combine this with the uncalibrated, wide prior, $\Delta A_{\rm u} = 3$ (chosen to be more uninformative in the more typical [-3,3] NLA prior range), in weighted quadrature at each redshift slice using Eqn.~\ref{eqn:uncal} to account for missing direct measurements,
\begin{equation}
    \Delta A_{\rm b}(z) = \sqrt{(1-f_{\rm u,b}(z))^2\Delta A_{\rm m}^2+f_{\rm u,b}(z)^2\Delta A_u^2} \; .
\end{equation}
We then produce the width of the per-bin Gaussian prior on alignment amplitude, $\Delta A_{\rm meb}$, as the mean of $\Delta A_{\rm b}(z)$, weighted by the tomographic redshift distribution of the sources, 
\begin{equation}
    \Delta A_{\rm b} = \frac{\int n_{\rm b}(z) \Delta A_{\rm b}(z) dz}{\int n_{\rm b}(z)dz} \; .
\end{equation}
See Fig.~\ref{fig:tailoredia} for an illustration of the redshift distributions, measurements, and fiducial per-bin priors resulting from this process. Note that the high-$z$ DESI measurement from \cite{Siegel2025_IA} is noisy and very sensitive to choice of scale -- consistent with no IA (see their Fig.~3) and not indicative of a trend towards negative alignment at high redshift. These priors are reported as Gaussians $(\mu,\sigma)$, with means $\mu$ and standard deviations $\sigma$ in Table~\ref{tab:priors} for the DES Y3 blue shear catalog made public in \cite{mccullough:2024_blue}. These Gaussian priors are truncated at [-3,3], appropriately conservative given the blue nature of the galaxy sample. The full method detailing tailored priors for intrinsic alignment, including aligning populations, will be released in the companion paper McCullough, Siegel et al. (in prep).

Fig.~\ref{fig:IAchoices} shows the marginalised $\Omega_{\rm m}$–$S_8$ posteriors constrained from the DES Y3 blue sample under different IA modelling choices. Using uninformed IA priors, with no tailored-IA information and flat $[-3,3]$ limits on each IA amplitude bin, yields $S_8 = 0.837^{+0.016}_{-0.022}$. Using tailored-IA priors improves the $S_8$ precision by a factor of 1.2, while shifting the mean by only $0.2\sigma$.  Analyzing the blue sample with no IA model at all, as was preferred in \cite{mccullough:2024_blue}, yields $S_8 = 0.828^{+0.013}_{-0.013}$, improving the precision on $S_8$ by a further factor of 1.2 and shifting $S_8$ with respect to tailored-IA by $0.2\sigma$. The goodness-of-fit for all choices of IA modeling remains relatively constant at $\chi^2_{\mathrm{min}}\sim380$, with the tailored choice being very slightly lower and the no-IA model having fewer effective degrees of freedom ($N_{\rm p}=3.02$), and the flat-per-bin-IA model having more ($N_{\rm p}=4.88$). With each choice using tailored priors on baryon feedback, we find $\chi^2_{\mathrm{red}}\approx 0.96$ for all models: flat-per-bin IA, tailored-IA, and no-IA. 

\begin{figure}
    \centering
    \includegraphics[width=0.8\linewidth]{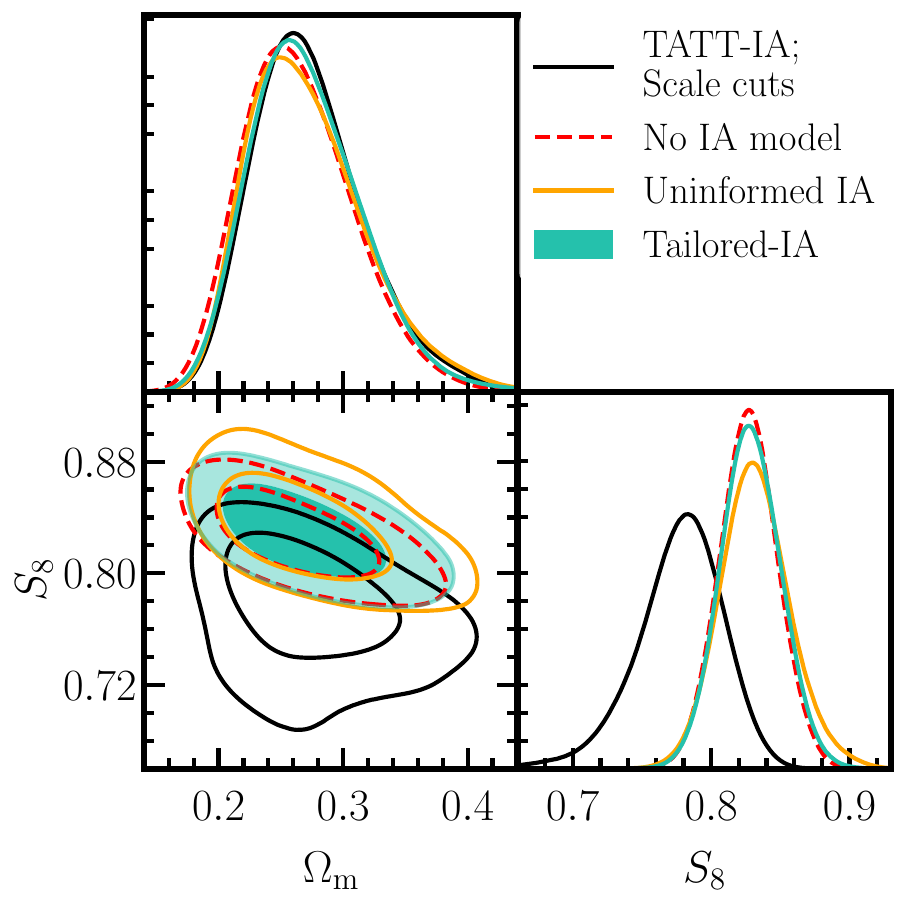}
    \caption{Marginalised $\Omega_{\rm m}-S_8$ posteriors constrained using the DES Y3 blue sample and different IA modelling choices: no IA model (red dashed), uninformed IA (orange) and tailored-IA (teal).  All variants are analyzed using calibrated feedback.  For reference, we also show the fiducial DES Y3 choices (scale cuts and TATT IA model; black).}
    \label{fig:IAchoices}
\end{figure}

\section{Cosmological results in context}\label{app:comparison}
Fig.~\ref{fig:s8_comp} compares our results with constraints on $S_8$ from other cosmological analyses. Historically, cosmic shear surveys have shown a small but persistent preference for lower values of $S_8$ relative to those from the primary CMB.  Compared to the fiducial DES astrophysical modelling, our `Tailored-IA and Calibrated feedback' strategy results in improved consistency with the CMB (Planck+ACT \citep{ACTDR6} at the 0.1$\sigma$ level) and $S_8$ measurements across redshift and from linear and non-linear scales.

\begin{figure}
    \centering
    \includegraphics[width=0.95\columnwidth]{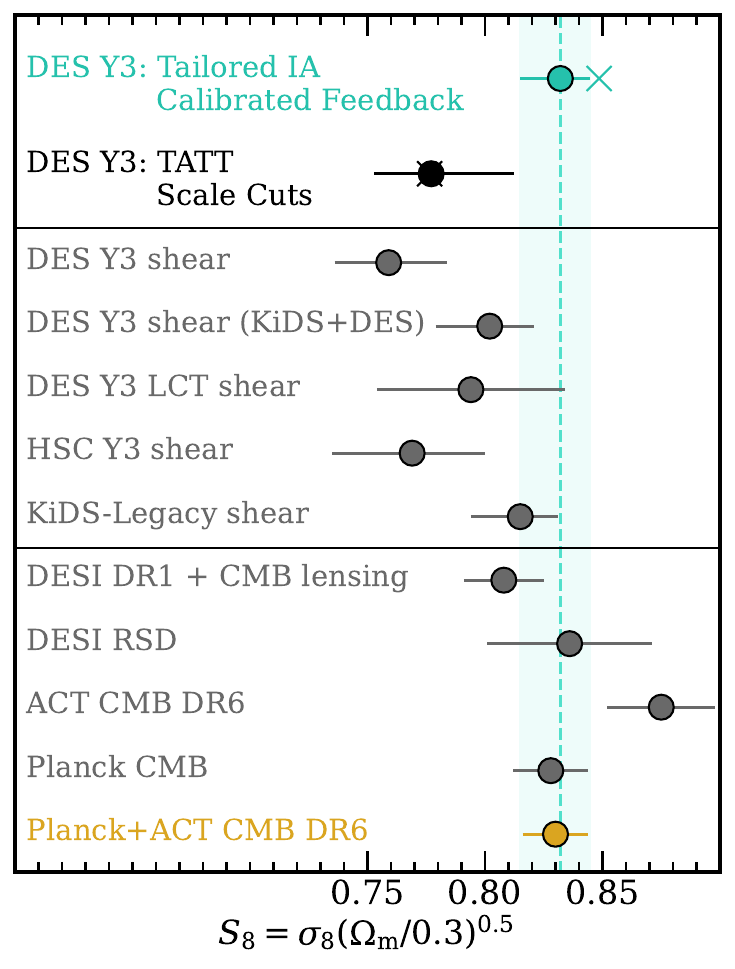}
    \caption{Marginalised $\Lambda$CDM constraints on $S_8$ from our reanalysis of DES Y3 using a Calibrated Feedback and Tailored-IA model (teal, top row) compared to the result using the fiducial Y3 model (TATT and Scale Cuts, second row); the results from each lensing survey (middle): DES Y3 \citep{amon_2022, SeccoSamuroff_2022}, DES Y3 with updated nonlinear modeling and scale-cuts in  the DES+KiDS hybrid pipeline \citep{KiDSDES}, DES Y3 analyzed with lensing counterterms (LCT) \citep{derose2025}, HSC Y3 \citep{li2023hyper}, and KiDS-Legacy \citep{wright2025}; and results from other cosmological observations and reanalyses: DESI RSD combined with CMB lensing \citep{Maus2025}, DESI RSD \citep{DESI-RSD}, and CMB (TTTEEE) from ACT DR6 \citep{ACTDR6}, Planck \citep{Planck2018} and their combination, including lensing \citep{ACTDR6}.  We show the mean and the 68\% confidence intervals, with the MAP estimate indicated by a cross. }
    \label{fig:s8_comp}
\end{figure}

\section{Consistency of X-ray and kSZ observations}\label{app:xraykszchoice}

Recent eROSITA measurements indicate lower halo gas mass fractions than many previous X-ray datasets \citep{Siegel:2025_kSZ}. Efforts to understand the factors driving these differences are ongoing (e.g. see discussion in Section~6.1 of \cite{Siegel2025_BC}).  In \cite{Siegel2025_BC} the suppression of the matter power spectrum is constrained separately by kSZ effect data \citep{ReidkSZ, Schaan:kSZ}, eROSITA X-ray \citep{Siegel:2025_kSZ} and HSC–XXL X-ray \citep{Akino2022} measurements (their Fig.~4).  In this Appendix, we assess the consistency of the cosmological constraints obtained when applying calibrated priors on the matter power spectrum suppression from the three datasets individually.

Fig.~\ref{fig:obs} shows the marginalized $\Omega_{\rm m}$–$S_8$ posteriors and the suppression of the matter power spectrum constraints obtained with feedback modelling calibrated separately on HSC–XXL, eROSITA and kSZ data, all analyzed with Tailored-IA.  We compare these to our headline `Tailored-IA, Calibrated feedback' result (teal), informed jointly by eROSITA and kSZ.  We constrain $S_8 = 0.820^{+0.016}_{-0.019}, 0.822^{+0.015}_{-0.019},0.834^{+0.017}_{-0.021}$ from the HSC–XXL, eROSITA and kSZ calibrated analyses respectively.  These are consistent within 1-$\sigma$. The suppression of the matter power spectrum constraints are also consistent at 1-$\sigma$ level. The HSC–XXL informed analysis shifts to $0.5\sigma$ lower $S_8$ than our `Tailored-IA, Calibrated feedback' approach using kSZ+eROSITA. This is anticipated given that the prior (see Fig.~4, \citet{Siegel2025_BC}) from the HSC–XXL analysis indicates a weaker mean suppression than the analyses with eROSITA and kSZ calibrated feedback.

\begin{figure}
    \centering
    \includegraphics[width=0.75\linewidth]{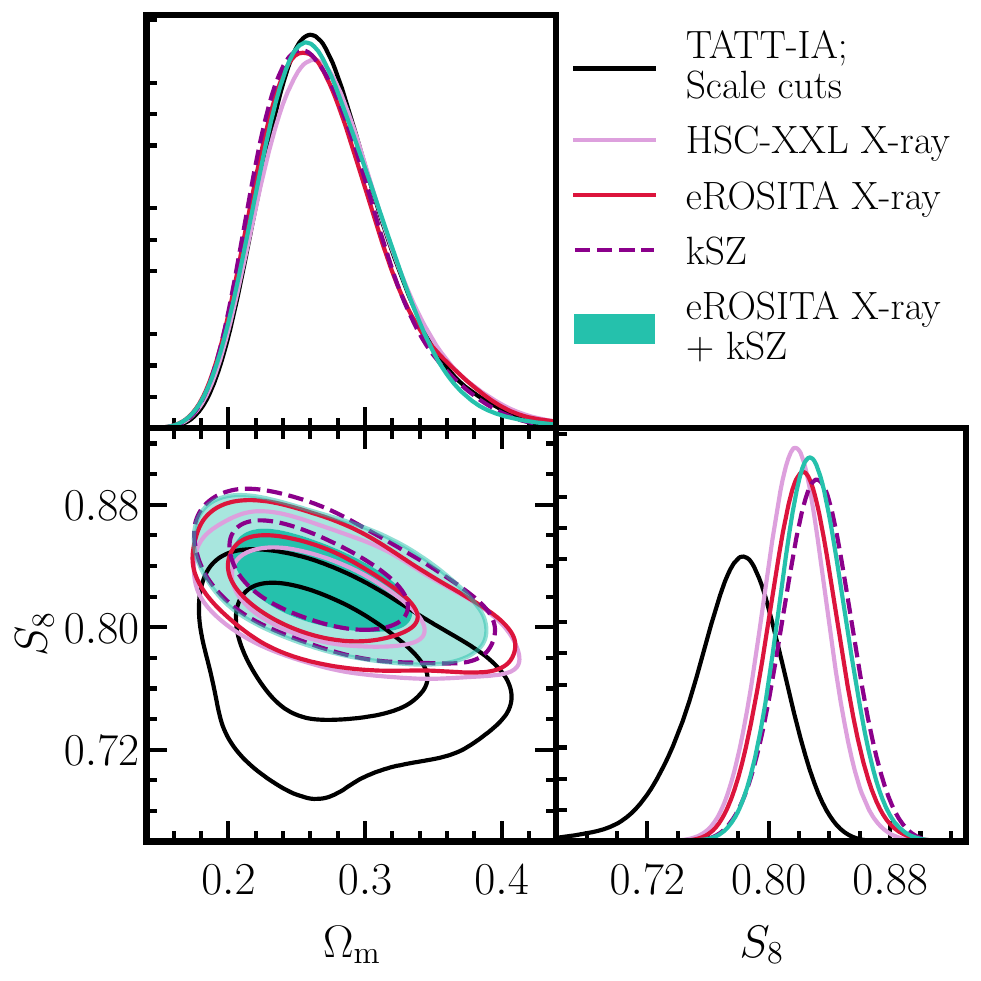}
    \includegraphics[width=0.75\linewidth]{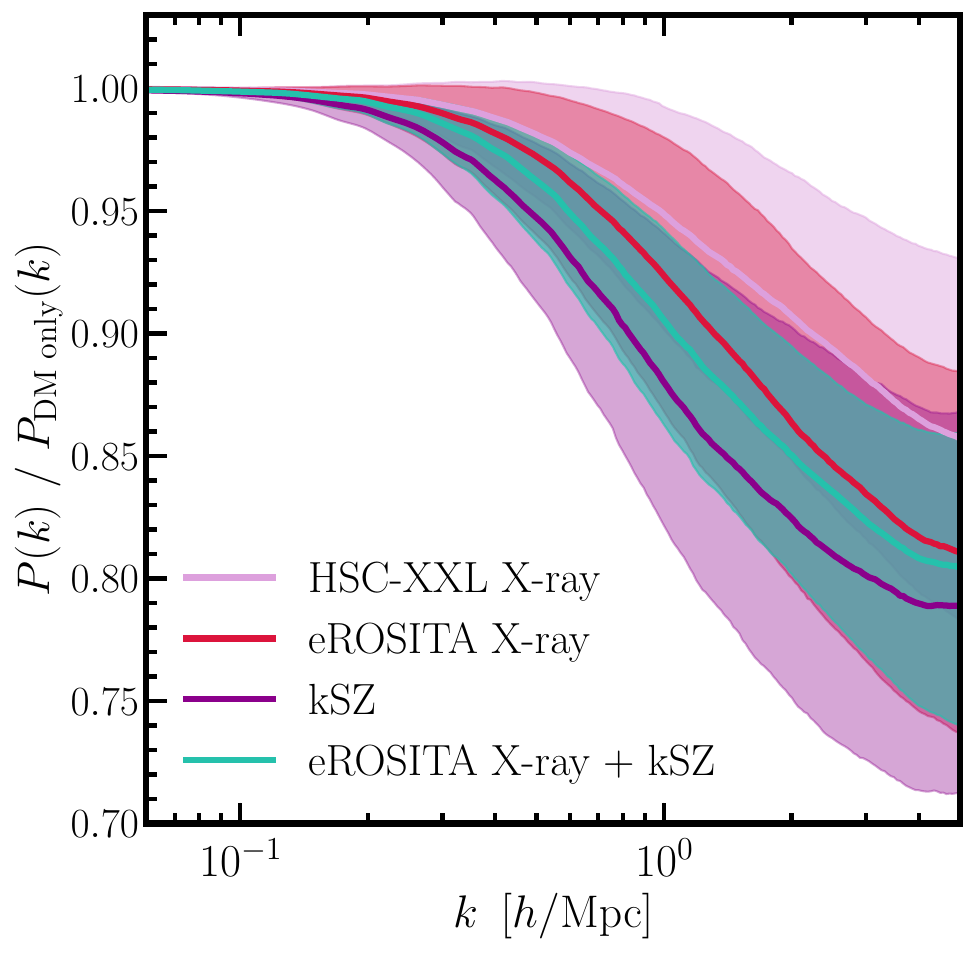}
    \caption{The consistency of analysing the DES Y3 blue sample with baryonic feedback calibrated individually on HSC–XXL \citep{Akino2022} (light purple), eROSITA \citep{Siegel:2025_kSZ} (red), and kSZ effect data \citep{ReidkSZ, Schaan:kSZ} (dark purple).  We also show the calibrated feedback model used throughout the text, informed by eROSITA and kSZ data (teal). \textit{Upper:} Marginalised $\Omega_{\rm m}-S_8$ posteriors.  \textit{Lower:} The suppression of the matter power spectrum.  All variants are analyzed with tailored-IA. }
    \label{fig:obs}
\end{figure}

\bibliographystyle{apsrev4-2_edit}
\bibliography{references}

\end{document}